\documentclass[a4paper,12pt,twoside]{article}

\usepackage{cite}
\usepackage[english]{babel}
\usepackage{rotating}
\usepackage{epsfig}
\usepackage[centertags]{amsmath}
\usepackage{amssymb,amscd,amsatdef}
\usepackage{array}
\usepackage{multirow}
\usepackage{supertabular}
\usepackage{dcolumn}
\usepackage{xspace}

\def\zero{{\scriptscriptstyle 0}}

\def\Z0{\ensuremath{Z^\zero}}

\def\pb{\,\text{pb}}

\def\R{{\scriptscriptstyle{\rm R}}}

\def\SU2U1{{\rm SU}(2)\times{\rm U}(1)}

\mathchardef\qsm=63
\mathchardef\pls=43
\mathchardef\mns=512
\mathchardef\plm=518
\mathchardef\eql=61
\mathchardef\smallleft=300
\mathchardef\smallright=301
\mathchardef\perslsh=47
\mathchardef\les=316
\mathchardef\gre=318
\mathchardef\leq=532
\mathchardef\grq=533
\chardef\usc=95
\chardef\til=126

\def\sqr#1#2#3{{\vcenter{\hrule height.#3ex\hbox{\vrule width.#2ex height#1ex
    \kern#1ex\vrule width.#3ex}\hrule height.#2ex}}}

\def\angleto{\vrule width.035em height2.1ex depth-.56ex\unskip\kern-.6ex\to}
\def\perchc#1{{\raise.4ex\hbox{$\mkern4mu#1{\it\perslsh}_
             {\mkern-5mu\scriptscriptstyle{{\rm o}\!{\rm o}}}^
             {\mkern-12.8mu\scriptscriptstyle{\rm o}}$}}}

\catcode`\@=11 
\def\parenbar{\mathpalette\p@renb@r}
\def\p@renb@r#1#2{\vbox{%
  \ifx#1\scriptscriptstyle \dimen@.7em\dimen@ii.2em\else
  \ifx#1\scriptstyle \dimen@.8em\dimen@ii.25em\else
  \dimen@1em\dimen@ii.4em\fi\fi \offinterlineskip
  \ialign{\hfill##\hfill\cr
    \vbox{\hrule width\dimen@ii}\cr
    \noalign{\vskip-.3ex}%
    \hbox to\dimen@{$\mathchar300\hfil\mathchar301$}\cr
    \noalign{\vskip-.3ex}%
    $#1#2$\cr}}}
\catcode`\@=12 

\newbox\struttbox
\setbox\struttbox=\hbox{\vrule height1.65ex depth.485ex width0pt}
\def\strutt{\relax\ifmmode\copy\struttbox\else\unhcopy\struttbox\fi}
\def\stru#1#2{\relax\ifmmode\hbox{\vrule height#1 depth#2 width0pt}
\else\vrule height#1 depth#2 width0pt\fi}

\def\ronum#1{\uppercase\expandafter{\romannumeral#1}}
\def\ronuml#1{\expandafter{\romannumeral#1}}

\DeclareMathAlphabet{\mathbf}{OT1}{cmr}{bx}{sl}

 {\end{list}}
 {\end{list}}
 {\end{list}}

\catcode`\@=11 
\newlength{\@fninsert}
\newlength{\@fnwidth}
\renewcommand{\@makefntext}[1]%
  {\noindent\makebox[\@fninsert][r]{\@makefnmark}\hfil%
  \parbox[t]{\@fnwidth}{#1}}
\catcode`\@=12 
\newlength{\localtextwidth}
\catcode`\@=11 
\newsavebox{\tmpbox}
\newlength{\@captionmargin}
\newlength{\@captionwidth}
\newlength{\@captionitemtextsep}
\renewcommand{\@makecaption}[2]%
  {%
   \vspace{10.pt}
   \setlength{\@captionwidth}{\localtextwidth}
   \addtolength{\@captionwidth}{-\@captionmargin}
   \sbox{\tmpbox}{{\bf #1:}{\rm #2}}%
   \ifthenelse{\lengthtest{\wd\tmpbox > \@captionwidth}}%
   {\centerline{\parbox[t]{\@captionwidth}%
   {\tolerance=2000\normalsize%
    {\bf #1:}\hspace{\@captionitemtextsep}{\rm #2}}}}%
   {\centerline{{\bf #1:}\kern1.em{\rm #2}}}}
\catcode`\@=12 
\catcode`\@=11 
\renewcommand\section{\@startsection{section}{1}{\z@}%
                                   {-3.5ex \@plus -1ex \@minus -.2ex}%
                                   {2.3ex \@plus.2ex}%
                                   {\normalfont\Large\bfseries}}
\renewcommand\subsection{\@startsection{subsection}{2}{\z@}%
                                   {-3.25ex\@plus -1ex \@minus -.2ex}%
                                   {1.5ex \@plus .2ex}%
                                   {\normalfont\large\bfseries}}
\renewcommand\subsubsection{\@startsection{subsubsection}{3}{\z@}%
                                   {-3.25ex\@plus -1ex \@minus -.2ex}%
                                   {1.5ex \@plus .2ex}%
                                   {\normalfont\large\bfseries}}
\renewcommand\paragraph{\@startsection{paragraph}{4}{\z@}%
                                   {3.25ex \@plus1ex \@minus.2ex}%
                                   {1.2ex \@plus .2ex}%
                                   {\normalfont\normalsize\bfseries}}
\catcode`\@=12 

\newsavebox{\sesbox}
\newlength{\seslen}

\def\etjet{E_T^{\rm jet}}
\def\etajet{\eta^{\rm jet}}
\def\phijet{\varphi^{\rm jet}}

\def\etcal{E_{T,{\rm cal}}^{\rm jet}}
\def\etacal{\eta_{\rm cal}^{\rm jet}}
\def\phical{\varphi_{\rm cal}^{\rm jet}}

\def\wcal{W_{\rm cal}}

\def\etar{-1<\etajet<2.5}

\def\etacr{-1<\etacal<2.5}

\def\qg2{$\q2>125$~\g2}

\def\wr{134 $<W<$ 277 GeV}

\def\q2{Q^2}
\def\pb1{pb$^{-1}$}
\def\gp{\gamma p}

\def\g2{GeV$^2$}
 
\def\rr1{R=1.0}
\def\r7{R=0.7}
\def\R71{R=0.7\ {\rm and}\ 1.0}

\def\mj{M^{\rm jj}}

\def\cost{\vert\cos\theta^*\vert}
\def\costcm{\vert\cos\theta^{\rm CM}\vert}

\def\smj{d\sigma/d\mj}

\def\schi{1/\sigma\ d\sigma/d\chi}
\def\rx{R_{\chi}}
\def\rxe{{\sigma(1<\chi<4)\over\sigma(4<\chi<9)}}

\def\kt{k_T}

\def\lq2{\log_{10}(\q2)}

\def\xo{x_{\gamma}^{\rm obs}}

\def\epr{e^+p\rightarrow e^+\ +\ {\rm jet}\ +\ {\rm jet}\ + \ {\rm X}}

\def\Journal#1#2#3#4{{#1}{#2} (#3) #4}
\def\NIM{{Nucl. Instr. and Meth.} A }
\def\NPB{{Nucl. Phys.} B }
\def\PLB{{Phys. Lett.}  B }
\def\PRL{{Phys. Rev. Lett.} }
\def\PRD{{Phys. Rev.} D }
\def\PRX{{Phys. Rev.} }
\def\PRE{{Phys. Rep.} }
\def\ZPC{{Z. Phys.} C }
\def\ZPX{{Z. Phys.} }
\def\EPC{{Eur. Phys. J.} C }
\def\CPC{{Comp. Phys. Comm.} }

\def\etal{{et al.}}

\def\rtr{< \hspace{-0.15cm} r_{\rm tracks} \hspace{-0.15cm} >}
\def\rtd{< \hspace{-0.15cm} r_{\rm dijet} \hspace{-0.15cm} >}
\def\rte{< \hspace{-0.15cm} r_{e/{\rm jet}} \hspace{-0.15cm} >}

\def\ptmis{p_T\hspace{-4.2mm}\slash\hspace{1.5mm}}

\def\wp{W^{\pm}}
\def\z0{Z^{0}}

\begin{document}
\selectlanguage{english}

\thispagestyle{empty}

\title{
\vspace{2cm}
\bf\LARGE High-mass dijet cross sections\\
\bf\LARGE in photoproduction at HERA
}
                    
\author{ZEUS Collaboration}

\date{ }
\maketitle

\vspace{-8.5cm}
\begin{flushleft}
\tt DESY-01-219 \\
December 2001 \\
\end{flushleft}
\vspace{+8cm}

\vfill
\centerline{\bf Abstract}
\vskip4.mm
\centerline{
  \begin{minipage}{15.cm}
    \noindent
Dijet differential cross sections for the reaction $\epr$ in the photoproduction
regime have been measured with the ZEUS detector at HERA using an integrated
luminosity of $42.7$~\pb1. The cross sections are given for photon-proton
centre-of-mass energies in the range \wr. The differential cross sections as a
function of the dijet mass, $\mj$, and of the dijet angular variables have been
measured for $47 < \mj < 160$~GeV  and compared to next-to-leading-order QCD
calculations. The dijet events in the region $75< \mj < 100$~GeV have been used to
derive a $95\%$~C.L. upper limit on the cross section for $\z0$ photoproduction of 
$\sigma_{e^+ p \rightarrow e^+ \z0 X} < 5.9$~pb. Upper limits on the
photoproduction of new heavy resonances decaying into two jets are also presented
for masses in the range between $60$~GeV and $155$~GeV.
  \end{minipage}
  }
\vfill

\thispagestyle{empty}
\pagenumbering{Roman}
\setcounter{page}{0}
\def\3{\ss}
\parindent0.cm
\parskip0.3cm plus0.05cm minus0.05cm
\newpage
\begin{center}
{\Large  The ZEUS Collaboration}
\end{center}
  S.~Chekanov, 
  M.~Derrick,                                                                                      
  D.~Krakauer,                                                                                     
  S.~Magill,                                                                                       
  B.~Musgrave,                                                                                     
  A.~Pellegrino,                                                                                   
  J.~Repond,                                                                                       
  R.~Yoshida\\                                                                                     
 {\it Argonne National Laboratory, Argonne, Illinois 60439-4815}~$^{n}$                            
\par \filbreak                                                                                     
  M.C.K.~Mattingly \\                                                                              
 {\it Andrews University, Berrien Springs, Michigan 49104-0380}                                    
\par \filbreak                                                                                     
  P.~Antonioli,                                                                                    
  G.~Bari,                                                                                         
  M.~Basile,                                                                                       
  L.~Bellagamba,                                                                                   
  D.~Boscherini,                                                                                   
  A.~Bruni,                                                                                        
  G.~Bruni,                                                                                        
  G.~Cara~Romeo,                                                                                   
  L.~Cifarelli,                                                                                    
  F.~Cindolo,                                                                                      
  A.~Contin,                                                                                       
  M.~Corradi,                                                                                      
  S.~De~Pasquale,                                                                                  
  P.~Giusti,                                                                                       
  G.~Iacobucci,                                                                                    
  G.~Levi,                                                                                         
  A.~Margotti,                                                                                     
  T.~Massam,                                                                                       
  R.~Nania,                                                                                        
  F.~Palmonari,                                                                                    
  A.~Pesci,                                                                                        
  G.~Sartorelli,                                                                                   
  A.~Zichichi  \\                                                                                  
  {\it University and INFN Bologna, Bologna, Italy}~$^{e}$
% ...
\newpage                                                                                           
% ...
                                                           %
                                                           %
\newpage   
                                                           %
                                                           %
\begin{tabular}[h]{rp{14cm}}                                                                       
$^{a}$ &  supported by the Natural Sciences and Engineering Research                               
          Council of Canada (NSERC) \\                                                             
$^{b}$ &  supported by the German Federal Ministry for Education and                               
          Research (BMBF), under contract numbers HZ1GUA 2, HZ1GUB 0, HZ1PDA 5, HZ1VFA 5\\         
$^{c}$ &  supported by the MINERVA Gesellschaft f\"ur Forschung GmbH, the                          
          Israel Science Foundation, the U.S.-Israel Binational Science                            
          Foundation, the Israel Ministry of Science and the Benozyio Center                       
          for High Energy Physics\\                                                                
$^{d}$ &  supported by the German-Israeli Foundation, the Israel Science                           
          Foundation, and by the Israel Ministry of Science\\                                      
$^{e}$ &  supported by the Italian National Institute for Nuclear Physics (INFN) \\                
$^{f}$ &  supported by the Japanese Ministry of Education, Science and                             
          Culture (the Monbusho) and its grants for Scientific Research\\                          
$^{g}$ &  supported by the Korean Ministry of Education and Korea Science                          
          and Engineering Foundation\\                                                             
$^{h}$ &  supported by the Netherlands Foundation for Research on Matter (FOM)\\                   
$^{i}$ &  supported by the Polish State Committee for Scientific Research,                         
          grant no. 115/E-343/SPUB-M/DESY/P-03/DZ 121/2001-2002\\                                  
$^{j}$ &  partially supported by the German Federal Ministry for Education                         
          and Research (BMBF)\\                                                                    
$^{k}$ &  supported by the Fund for Fundamental Research of Russian Ministry                       
          for Science and Edu\-cation and by the German Federal Ministry for                       
          Education and Research (BMBF)\\                                                          
$^{l}$ &  supported by the Spanish Ministry of Education and Science                               
          through funds provided by CICYT\\                                                        
$^{m}$ &  supported by the Particle Physics and Astronomy Research Council, UK\\                   
$^{n}$ &  supported by the US Department of Energy\\                                               
$^{o}$ &  supported by the US National Science Foundation                                          
\end{tabular}                                                                                      
                                                           %
                                                           %
\newpage
\parindent0.cm
\parskip         0.2cm plus0.05cm minus0.05cm
\pagenumbering{arabic}
\setcounter{page}{1}
\raggedbottom
\normalsize

\section{Introduction}
\label{secintro}
In hadronic interactions, the distributions of the dijet mass, $\mj$, and of the
angle, $\theta^{\rm CM}$, between the jet-jet axis and the beam direction in the
dijet centre-of-mass system, provide a test of QCD. In addition, they are sensitive
to the presence of new particles that decay into two jets. At HERA, these tests of
QCD and searches for new particles can be made in photoproduction via
$ep$ scattering at $\q2\approx 0$, where $\q2$ is the virtuality of the exchanged
photon. Two types of QCD processes contribute to jet production at leading order
(LO) in photoproduction \cite{owens,drees}: either the photon interacts directly
with a parton in the proton (the direct process) or the photon acts as a source of
partons, one of which interacts with a parton in the proton (the resolved process).
Jet photoproduction thereby provides a means to study the dynamics of both
photon-parton and parton-parton interactions.

At high $\mj$ values, the theoretical uncertainties due to hadronisation,
multipartonic interactions and the limited knowledge of the photon parton
densities are reduced. The dynamics of dijet production for hadron-induced
processes has been investigated in detail in $p\bar{p}$ collisions at
centre-of-mass energies of $\sqrt{s}=630$~GeV \cite{dijuas,ua1def} and 
$\sqrt{s}=1800$~GeV \cite{dijtev}. Next-to-leading-order (NLO) QCD predictions are
in good agreement with the measured dijet mass and angular distributions up to
$\mj \approx 1000$~GeV. Previous measurements of the dijet mass and angular
distributions \cite{dijuas,ua1def,dijtev,angular} made use of cone algorithms for
the identification of the jets. The $\kt$ cluster algorithm \cite{kt,seymour} is
used here, which allows a direct application of the theoretical jet algorithm to
the data. The selection of high $\mj$ values permits a precise test of the
description of the dynamics of dijet photoproduction to smaller distances than
previously studied in photoproduction at HERA \cite{angular}.

Heavy particles that decay into two jets would show up as an enhancement in
the dijet mass distribution. In particular, this mass distribution is
sensitive to the production of the electroweak gauge bosons $\wp$ and $\z0$.
New heavy particles decaying into two jets may also be identified by deviations
from the predictions of QCD in the $\costcm$ distribution; in QCD, the $\costcm$
distribution is peaked at unity, whereas many sources of new physics produce more
isotropic angular distributions.

In this paper, measurements of the dijet differential cross-sections $\smj$
and $\schi$, where $\chi \equiv (1+\cost)/(1-\cost)$, are presented as a
function of $\mj$. The variable $\theta^*$ is defined by using the
pseudorapidities\footnote{The pseudorapidity is defined as
$\eta=-\ln\left(\tan\frac{\theta}{2}\right)$, where the polar angle, $\theta$,
is measured with respect to the proton beam direction.} of the two jets with highest
transverse energy in the event, $\cos{\theta^*}\equiv \tanh
\frac{1}{2}(\eta^{\rm jet1}-\eta^{\rm jet2})$, and coincides with $\theta^{\rm CM}$
for the case of $2 \rightarrow 2$ massless parton scattering. To
increase the sensitivity to direct processes, measurements have also been made for
dijet events in which the fraction of the photon's momentum participating in the
production of the dijet system is greater than $0.75$. The NLO QCD calculations 
\cite{klasen,harris,frixione,aurenche} are compared to the present
measurements and upper limits are set on the inclusive cross sections for the
photoproduction of $\wp$ and $\z0$ bosons. A search has also been carried out
for the photoproduction of new heavy resonances decaying into two jets.

\section{Perturbative QCD calculations}
\label{secqcd}
Perturbative QCD (pQCD) calculations of dijet cross sections in photoproduction
can be written as a convolution of the subprocess cross section with
the parton distribution functions (PDFs) of the photon and proton:
\begin{equation}
 d\sigma_{e p \rightarrow e\ {\rm jet}\ {\rm jet} X} = 
 \displaystyle\sum_{a,b} \int_0^1 dy f_{\gamma/e}(y) 
   \int_0^1 dx_{\gamma} f_{a/\gamma}(x_{\gamma},\mu^2_{F\gamma}) 
   \int_0^1 dx_{p} f_{b/p}(x_{p},\mu^2_{Fp}) \; 
   d\hat{\sigma}_{ab \rightarrow {\rm jet}\ {\rm jet}}(\mu_R), \nonumber
\end{equation}
where $y$, $x_{\gamma}$ and $x_p$ are the longitudinal momentum fractions of the
quasi-real photon emitted by the positron, the parton $a$ in the photon and the
parton $b$ in the proton, respectively; $f_{\gamma/e}$ is the flux of photons in
the positron and is usually estimated with the Weizs\"acker-Williams approximation
\cite{wwa}; $f_{a/\gamma}$ ($f_{b/p}$) represents the PDF of parton $a$ ($b$) in
the photon (proton) $-$ in the case of direct processes, $a$ is a $\gamma$ and 
$f_{a/\gamma}(x_{\gamma},\mu^2_{F\gamma})$ is given by $\delta(1-x_{\gamma})$;
the factorisation scale in the photon (proton) is denoted by $\mu_{F\gamma}$
($\mu_{Fp}$); $\mu_R$ represents the renormalisation scale; and the subprocess
cross section, $d\hat{\sigma}_{ab\rightarrow {\rm jet}\ {\rm jet}}$, describes
the short-distance structure of the interaction.

A wealth of data from fixed-target and collider experiments has made possible an
accurate determination of the proton PDFs. In the case of the photon,
experimental information on the quark densities is available from measurements
of $F_2^{\gamma}$ in $e^+e^-$ collisions, while the gluon density remains poorly
constrained. At high values of the dijet mass, the direct processes dominate and
thus the sensitivity to the photon PDFs is reduced.

The subprocess cross section is calculable in pQCD at each order. Recently, NLO
QCD calculations in photoproduction \cite{klasen,harris,frixione,aurenche} have
become available. In NLO QCD, the dependence of the calculations on the
renormalisation and factorisation scales is reduced compared to LO.
The results from the different NLO calculations have been compared
and found to be in agreement within $5\%$ in most of the phase-space region
studied \cite{khf}.

\section{Experimental conditions}
\label{secsetup}
During 1995-1997, HERA operated with protons of energy $E_p=820$~GeV and
positrons of energy $E_e=27.5$~GeV. The data sample used in this analysis
corresponds to an integrated luminosity of
$42.7\pm 0.7$~\pb1. The ZEUS detector is described in detail elsewhere
\cite{sigtot,status}. The main subdetectors used in the present analysis are the
central tracking detector (CTD) positioned in a 1.43~T solenoidal magnetic field
and the uranium-scintillator sampling calorimeter (CAL). 

Tracking information is provided by the CTD \cite{zeusctd}, which is used to
reconstruct the momenta of tracks in the polar-angle\footnote{The ZEUS coordinate
system is a right-handed Cartesian system, with the $Z$ axis pointing in the
proton beam direction, referred to as the ``forward direction'', and the $X$
axis pointing left towards the centre of HERA. The coordinate origin is at the
nominal interaction point.} region $15^\circ < \theta < 164^\circ$. The relative
transverse momentum, $p_T$, resolution for full-length tracks can be
parameterised as $\sigma(p_T)/p_T=0.0058\ p_T \oplus 0.0065\oplus 0.0014/p_T$,
with $p_T$ in GeV. The tracking system was used to establish an interaction
vertex with a typical resolution along (transverse to) the beam direction of
0.4~(0.1)~cm and to cross-check the energy scale of the CAL.

The CAL \cite{zeuscal} covers $99.7\%$ of the total solid angle. The smallest
subdivision of the CAL is called a cell. Energy
deposits in the CAL were used in the jet finding and to measure jet
energies. Under test-beam conditions, the CAL single-particle relative energy
resolutions were $0.18/\sqrt{E\,({\rm GeV})}$ for electrons and
$0.35/\sqrt{E\,({\rm GeV})}$ for hadrons. Jet energies were corrected (see
Section~\ref{seccorr}) for the energy lost in inactive material in front of the CAL.

The luminosity was measured using the Bethe-Heitler reaction $e^+ p
\rightarrow e^+\gamma p$~\cite{zeuslumi}. The resulting small-angle energetic
photons were measured by the luminosity monitor, a lead-scintillator
calorimeter placed in the HERA tunnel at $Z=-107$~m.

\section{Data selection and jet search}
\label{secsel}
A three-level trigger was used to select events online \cite{status,jetshapes}.
Events from collisions between quasi-real photons and protons were selected
offline using criteria similar to those described in an earlier ZEUS publication
\cite{highet}. The main steps are briefly discussed here. After requiring
a reconstructed event vertex consistent with the nominal interaction position
and cuts based on the tracking information, the contamination from beam-gas
interactions, cosmic-ray showers and beam-halo muons was negligible. Charged
current deep inelastic scattering (DIS) events were rejected by requiring the
total missing transverse momentum, $\ptmis$, to be small compared to the total
transverse energy, $E^{\rm tot}_T$: \mbox{$\ptmis/\sqrt{E^{\rm tot}_T}<2\ \sqrt{\rm GeV}$}.
Neutral current DIS events with an identified scattered-positron candidate
\cite{sinistra} in the CAL were removed from the sample using the method described
in an earlier publication \cite{zedir}. The remaining background from neutral
current DIS events was estimated by Monte Carlo techniques to be below $0.5\%$.
The selected sample consisted of events from $e^+p$ interactions with
$\q2\lesssim 1$ \g2\ and a median $\q2\approx 10^{-3}$~\g2. The events were
restricted to $\gamma p$ centre-of-mass energies in the range \mbox{\wr}, as described
in Section~\ref{secrec}.

The $\kt$ cluster algorithm \cite{kt} was used in the longitudinally invariant
inclusive mode \cite{ellissoper} to reconstruct jets in the hadronic final state
from the energy deposits in the CAL cells. The jet search was performed in the
pseudorapidity ($\eta$)-azimuth ($\varphi$) plane of the laboratory frame. The
jet variables were defined according to the Snowmass convention \cite{snow}. The
jets reconstructed from the CAL cell energies are called calorimetric jets and
the variables associated with them are denoted by $\etcal$, $\etacal$ and
$\phical$. There were 64708~events selected with at least two jets satisfying
$\etcal>10$~GeV and $\etacr$.

\section{Monte Carlo simulation}
\label{secmc}
The Monte Carlo (MC) programs PYTHIA~5.7~\cite{pythia} and HERWIG~5.9~\cite{herwig}
were used to generate photoproduction events for resolved and direct processes.
Events were generated using GRV-HO \cite{grv} for the photon PDFs and MRSA
\cite{mrsa} for the proton PDFs. To study the dependence of the acceptance
corrections on the choice of photon PDFs, the LAC1 parameterisations \cite{lac1}
were used. In both generators, the partonic processes are simulated using LO matrix
elements, with the inclusion of initial- and final-state parton showers.
Fragmentation into hadrons is performed using the LUND \cite{lund} string model as
implemented in JETSET \cite{jetset} in the case of PYTHIA, and the cluster model
\cite{webber} in the case of HERWIG. For the measurements presented in this paper,
the events generated using the PYTHIA and HERWIG programs were used for calculating
jet-energy corrections and correcting for detector and acceptance effects. The
corrections provided by PYTHIA were used as default values and those given by
HERWIG were used to estimate the systematic uncertainties coming from the
treatment of the parton shower and hadronisation.

Photoproduction of the electroweak gauge bosons $\wp$ and $\z0$ was simulated
with PYTHIA using the lowest-order processes $q \bar{q} \rightarrow \z0$ and
$q \bar{q}^{\prime} \rightarrow \wp$. Events were generated, including initial-
and final-state parton showers, using GRV-HO for the photon and MRSA for the proton
PDFs.

To model the photoproduction of narrow heavy resonances decaying to dijets,
MC events were simulated with PYTHIA using the LO process
$q \bar{q} \rightarrow Z^{\prime}$ and the same photon and proton PDFs as above.
Samples of events were generated for each value of the mass from $60$~GeV
up to $155$~GeV in $5$~GeV intervals. The vector and axial couplings
to quarks and leptons were set equal to those of the Standard Model $\z0$,
resulting in a width that increases linearly with mass. The cut $\cost < 0.6$
was used to minimise the sensitivity of the acceptance corrections to the
{\it  a priori} unknown decay angular distribution of a new narrow state.

All generated events were passed through the ZEUS detector and trigger
simulation programs based on GEANT~3.13 \cite{geant}. They were reconstructed and
analysed by the same program chain as the data. The jet search was
performed using the energy measured in the CAL cells in the same way as for
the data. The same jet algorithm was also applied to the final-state
particles; the jets found in this way are referred to as hadronic jets.

\section{Jet energy correction and selection}
\label{seccorr}

\subsection{Jet hadron-level corrections}
The comparison of the reconstructed jet variables for the hadronic and 
the calorimetric jets in simulated events showed that no correction was needed
for $\etajet$ and $\phijet$ ($\etajet\simeq\etacal$ and
$\phijet \simeq \phical$). However, the transverse energy of the calorimetric
jet underestimated that of the corresponding hadronic jet by an average
of $\sim 15\%$, with an r.m.s. of $\sim 10\%$. The transverse energy
corrections to calorimetric jets, averaged over the azimuthal angle, were first
determined using the MC events. These corrections were constructed as factors,
$C(\etcal,\etacal)$, which, when multiplied by $\etcal$, provide the corrected
transverse energies of the jets, $\etjet$.

\subsection{Jet energy-scale corrections}
Further corrections to the jet transverse energy were developed to account for 
differences in the jet energy scale between data and simulations.
This procedure relies on good understanding of the performance of the
CTD track reconstruction in the selected region (see below).

 The response of the CAL to jets was investigated by using the following
procedure \cite{highet,thesis}. In the central region, $|\etajet|<1$, the
multiplicity distribution and the $p_T$ spectrum of charged particles
associated with the calorimetric jets were compared for data and MC
samples using the reconstructed tracks. The tracks were required to be
in  the ranges $|\eta^{\rm track}|<1.5$ and $p^{\rm track}_T>300$~MeV, where
$p^{\rm track}_T$ is the transverse momentum of the track with respect to the beam
axis and $\eta^{\rm track}$ is the track pseudorapidity. Tracks were associated
with a calorimetric jet when the extrapolated track trajectory reached the CAL
within a cone of one unit radius in the $\eta$-$\varphi$ plane concentric with the
calorimetric-jet axis. PYTHIA gives a good description of all the measured
distributions. In this $\etajet$ region, the momenta of the tracks in the
calorimetric jet were used to determine the total transverse energy carried by the
charged particles, $E^{\rm jet}_{T,{\rm tracks}}$, assuming zero mass for
all tracks. Then, the ratio
$r_{\rm tracks}\equiv E^{\rm jet}_{T,{\rm tracks}}/\etcal$ was calculated and
the distributions of this ratio for the dijet sample in data and MC generated
events were compared. The mean value of the distribution in $r_{\rm tracks}$ was
determined as a function of $\etajet$ for data ($\rtr_{\rm data}$) and MC events
($\rtr_{\rm MC}$). Differences between data and MC simulation of less than $1\%$
were observed from the examination of the quantity
$(\rtr_{\rm data}/\rtr_{\rm MC})-1$. The transverse energies of the calorimetric
jets in the data were then modified as a function of $\etajet$ to correct for
these differences.

 In the forward region, $1<\etajet<2.5$, the energy scale of the jets was studied
using the transverse-energy imbalance in dijet events with one jet in the
central region and the other in the forward region. The distributions of the
ratio $r_{\rm dijet}\equiv\etcal$(forward jet)$/\etcal$(central jet) in data and
the MC sample were compared. Differences between data and MC simulation of less
than $2\%$ were observed from the examination of the quantity 
\mbox{$(\rtd_{\rm data}/\rtd_{\rm MC})-1$}. The transverse energies of the forward
calorimetric jets in the data were then modified as a function of $\etajet$ to
correct for these differences. The widths of the distributions for $r_{\rm tracks}$
and $r_{\rm dijet}$ are reasonably well described by the PYTHIA MC simulation,
giving confidence that the resolution in the energy of the jets is correctly
described.

 The accuracy of the jet energy corrections was investigated by using data and
simulations of neutral current DIS events at large $Q^2$, where the transverse energy
of the jet or jets is balanced by that of the scattered positron. The uncertainty
in the absolute CAL energy calibration for the scattered positrons was $1\%$
for positron energies above $15$~GeV~\cite{f2zeus01}. Jets were reconstructed in the
laboratory frame using the algorithm described in Section~\ref{secsel}. The
jet energy corrections described above were applied to both data and MC events. 
Then, the ratio $r_{e/{\rm jet}} \equiv E^{e}_T/\etjet$, where 
$E^{e}_T$ is the positron transverse energy, was formed and the distributions
of this ratio for the inclusive jet sample in DIS data and MC simulation
were compared. The mean value of the distribution of $r_{e/{\rm jet}}$ was
determined as a function of $\etajet$ in the range $\etar$ and of $\etjet$
in the range $14 < \etjet < 90$~GeV for data ($\rte_{\rm data}$) and MC events
($\rte_{\rm MC}$). Inspection of the quantity 
\mbox{$|(\rte_{\rm data}/$}\mbox{$\rte_{\rm MC})-1|$}
showed that the differences between data and MC simulation were smaller than
$1\%$. This variation was therefore included as a systematic uncertainty in the
present analysis.

After these further corrections to the jet transverse energy, events with
at least two jets satisfying $\etjet>14$ GeV and $\etar$ were used to measure the
dijet cross sections presented in Section~\ref{secres}.

\section{Reconstruction of jet and kinematic variables}
\label{secrec}
The invariant mass of the two jets with highest $\etjet$ in the event
was reconstructed using the formula
\begin{equation}
 \mj = \sqrt{2 E_T^{\rm jet1} E_T^{\rm jet2}
                [ \cosh{(\eta^{\rm jet1}-\eta^{\rm jet2})} -
                  \cos{(\varphi^{\rm jet1}-\varphi^{\rm jet2})}]}. \nonumber
\end{equation}
Only the absolute value of $\cos{\theta^*}$ can be determined because the
originating parton cannot be identified. For $\mj > 47$~GeV and $\cost < 0.8$,
the average relative resolutions in $\mj$ and $\chi$ were $8\%$ and $7\%$,
respectively; the average resolution in $\cost$ was $0.02$.

The Jacquet-Blondel method \cite{jacblo}, applied to photoproduction
events \cite{jorges}, was used to estimate $W$ from the energies measured in
the CAL cells, $\wcal$. Due to the energy lost in the inactive material in
front of the CAL and to particles lost in the rear beampipe, $\wcal$
underestimates $W$ by $\sim 10\%$, with an r.m.s. of $\sim 5\%$ \cite{jetshapes}.
This effect was corrected for using the MC simulation.

The fraction of the photon momentum participating in the production of the
two jets with highest $\etjet$ is defined \cite{zedir,zedij95} as
\begin{equation}
 \label{eqxgam}
 \xo = \frac{1}{2 y E_e}(E_T^{\rm jet1} e^{-\eta^{\rm jet1}}+
                       E_T^{\rm jet2} e^{-\eta^{\rm jet2}}),
\end{equation}
where the variable $y$ is given by $y=W^2/s$. The LO direct and resolved
processes populate different regions in $\xo$, with the direct processes
concentrated at high values of $\xo$. The variable $\xo$ was reconstructed via
the above formula using the calorimetric-jet transverse energies and
$y_{cal}={\wcal^2}/s$, since many systematic uncertainties in the measurement of
energy by the CAL cancel out event by event in the ratio of Eq.~(\ref{eqxgam}).
The average resolution in $\xo$ was $0.05$ for $\mj > 47$~GeV and $\cost < 0.8$.

\section{Acceptance corrections and systematic uncertainties}
\label{secacc}

The PYTHIA MC event samples of resolved and direct processes were used to
compute the acceptance corrections to the dijet distributions. These correction
factors also take into account the efficiency of the trigger, the selection
criteria and the purity and efficiency of the jet reconstruction. The
contributions from direct and resolved processes in the MC models were added
according to a fit to the uncorrected $\xo$ distribution in the data. A good
description of the $\mj$, $\cost$ and $\chi$ data distributions was given both by
PYTHIA and HERWIG. The differential dijet cross sections were obtained by applying
bin-by-bin corrections to the measured distributions. The bin-by-bin correction
factors differed from unity typically by less than $10\%$.

A detailed study of the sources of systematic uncertainty was performed. This study
includes (a typical contribution for each item to the cross-section uncertainty is
indicated in parentheses):

\begin{itemize}
 \item using the HERWIG generator to evaluate the correction factors for the
      observed dijet distributions ($5\%$);
 \item varying the cuts used to select the data while maintaining agreement between
      data and MC simulations ($2\%$);
 \item adding the contributions from direct and resolved processes according
      to the default cross sections as predicted by PYTHIA ($2\%$);
 \item using the LAC1 parameterisations of the photon PDFs for the PYTHIA MC
        samples ($2\%$).
\end{itemize}

The effects of uncertainties in the simulation of the trigger were
negligible. All the above systematic uncertainties were added in quadrature to
the statistical uncertainties.

The absolute energy scale of the calorimetric jets in simulated events was
varied by $\pm 1\%$. The effect of this variation on the dijet cross sections
was approximately $\mp 5\%$. This uncertainty is highly correlated between 
measurements in different bins. It is shown as a shaded band in Fig.~\ref{fig1}.
In Figs.~\ref{fig2} to \ref{fig4}, it has been added in quadrature to the
other systematic uncertainties. An overall normalisation uncertainty of $1.6\%$,
arising from the luminosity determination, is not included.

\section{High-mass dijet differential cross sections}
\label{secres}
Using the selected data sample of dijet events, the differential dijet
cross sections were measured in the kinematic region defined by
$\q2 \; \leq \; 1$~\g2\ and \wr. The dijet variables and cross sections were
calculated using the two highest-$\etjet$ hadronic jets with $\etjet > 14$~GeV and
$\etar$. For a given $W$, events at high $\cost$ have smaller scattering angles
and thus lower $\etjet$. In order to study the $\cost$ and $\mj$ distributions
without bias from the $\etjet$ requirement, the cuts $\mj > 47$~GeV and
$\cost <0.8$ were applied\footnote{The application of a lower cut on $\mj$ which is
not close to the threshold avoids the infrared sensitivity \cite{frixione} that
otherwise would affect NLO QCD calculations for dijet production with symmetric
cuts in the transverse energy of each of the two jets.}.

The cross-section $\smj$, measured in the $\mj$ range between 47 and 160~GeV and
integrated over $\cost<0.8$, is presented in Fig.~\ref{fig1}(a). The data points are
located at the mean of each $\mj$ bin. The measured $\smj$ exhibits a steep
fall-off over 3 orders of magnitude in the $\mj$ range considered.

The cross-section $\smj$ was also measured for the region $\xo > 0.75$. The results
are also presented in Fig.~\ref{fig1}(a) and exhibit a dependence on $\mj$ similar
to that of the measurements integrated over the full range in $\xo$. The fraction of
dijet events with $\xo >0.75$ increases with $\mj$ from $57\%$ at
$\mj =50$~GeV to $98\%$ at $\mj =139$~GeV.

To study the angular distribution as a function of the dijet mass, the
normalised dijet cross section, $\schi$, was measured for $1 ~\leq ~\chi~ \leq ~9$
in four regions of $\mj$; $\sigma$ is the dijet cross section integrated over 
the entire $\chi$ range in each $\mj$ region. The normalised cross
section has the advantage that the experimental and theoretical uncertainties
are reduced, while allowing a precise test of the shape of the distribution
in the calculations. The results integrated over the
full range in $\xo$ are presented in Fig.~\ref{fig2} and those in the region
$\xo >0.75$ are shown in Fig.~\ref{fig3}. The measured $\schi$ decreases with
increasing $\chi$ in all the $\mj$ regions studied.

\subsection{Perturbative QCD predictions}
\label{secnlo}
LO and NLO QCD calculations \cite{klasen} are compared to the measurements
in Figs.~\ref{fig1} to \ref{fig3}. The calculations were performed using the
GRV \cite{grv} and CTEQ4 \cite{cteq4} parameterisations of the photon and proton
PDFs, respectively. The renormalisation and factorisation scales were chosen to
be the highest $\etjet$; $\alpha_s$ was calculated at the two-loop level using
$\Lambda^{(5)}_{\overline{\rm MS}}=202$~MeV, which corresponds to 
$\alpha_s(M_{Z})=0.116$. The calculations included only QCD hard-scattering 
processes and thus electroweak corrections such as $q\bar q\rightarrow q\bar q$
via $\gamma^*/\z0$ exchange or $q\bar q^{\prime}\rightarrow q\bar q^{\prime}$
via $\wp$ exchange were not included. 

The uncertainty on the NLO calculations due to the absence of higher-order terms
was estimated by varying the choice of the renormalisation and factorisation
scales between $\etjet/2$ and $2\etjet$ and amounts to less than $15\%$ in the
case of $\smj$. This uncertainty affects mainly the normalisation of the
predictions and is therefore reduced for $\schi$, for which it typically amounts
to less than $5\%$. The uncertainty on the NLO calculations due to that on the
gluon density of the proton was estimated by comparing with calculations based
on the CTEQ4HJ \cite{cteq4} parameterisations and amounts to less than $6\%$ for
$\smj$. The uncertainty due to the choice of photon PDFs was estimated by
using the AFG \cite{afg} parameterisations. The choice of parameterisation of the
photon PDFs affects mainly the normalisation of the calculation. The normalisation
for the case of AFG is smaller by approximately $10\%$ than that based on GRV.
All the above theoretical uncertainties were added in quadrature.

NLO QCD calculations refer to jets of partons, whereas the measurements refer
to jets of hadrons. An estimate of the effects of hadronisation was
obtained by comparing the cross sections for jets of hadrons and jets of
partons calculated with the PYTHIA program. The corrections for hadronisation
effects were found to be within $\pm 5\%$ of unity both for $\smj$ and $\schi$;
they therefore were neglected in the comparison of pQCD calculations to the
measurements.

\subsection{Dijet mass distribution}

\begin{figure}
\setlength{\unitlength}{1.0cm}
\begin{picture} (18.0,17.5)
\put (1.0,5.75){\epsfig{figure=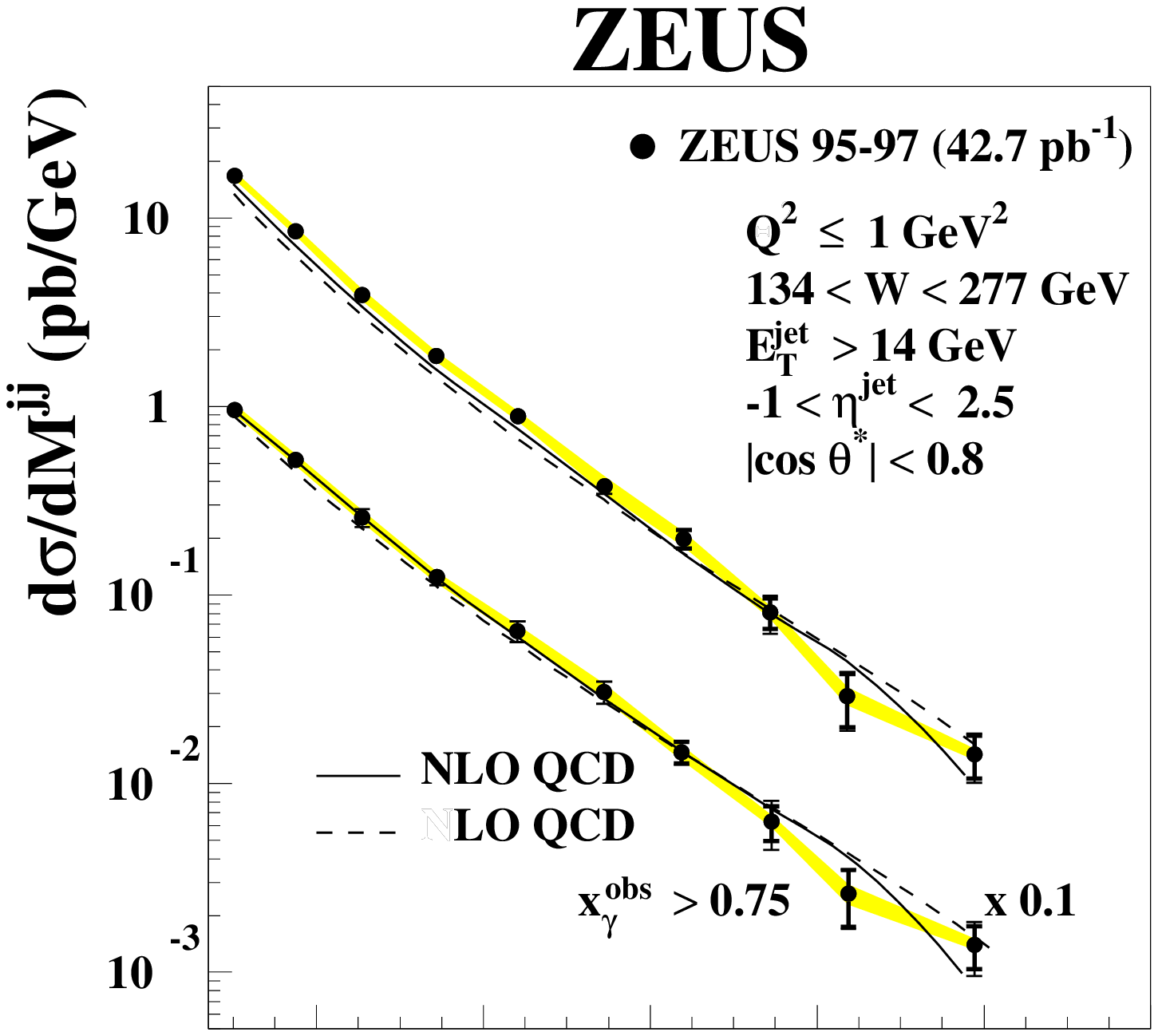,width=15cm}}
\put (1.0,-3.25){\epsfig{figure=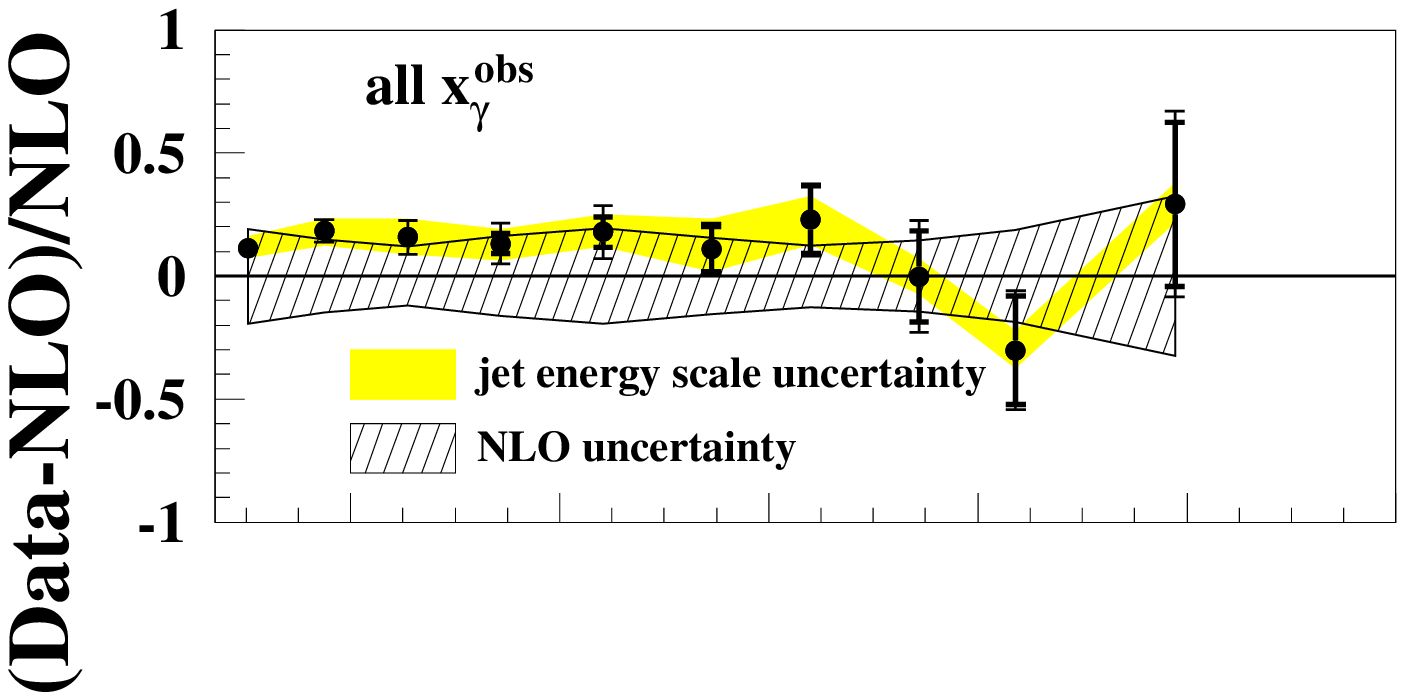,width=15cm}}
\put (1.0,-7.00){\epsfig{figure=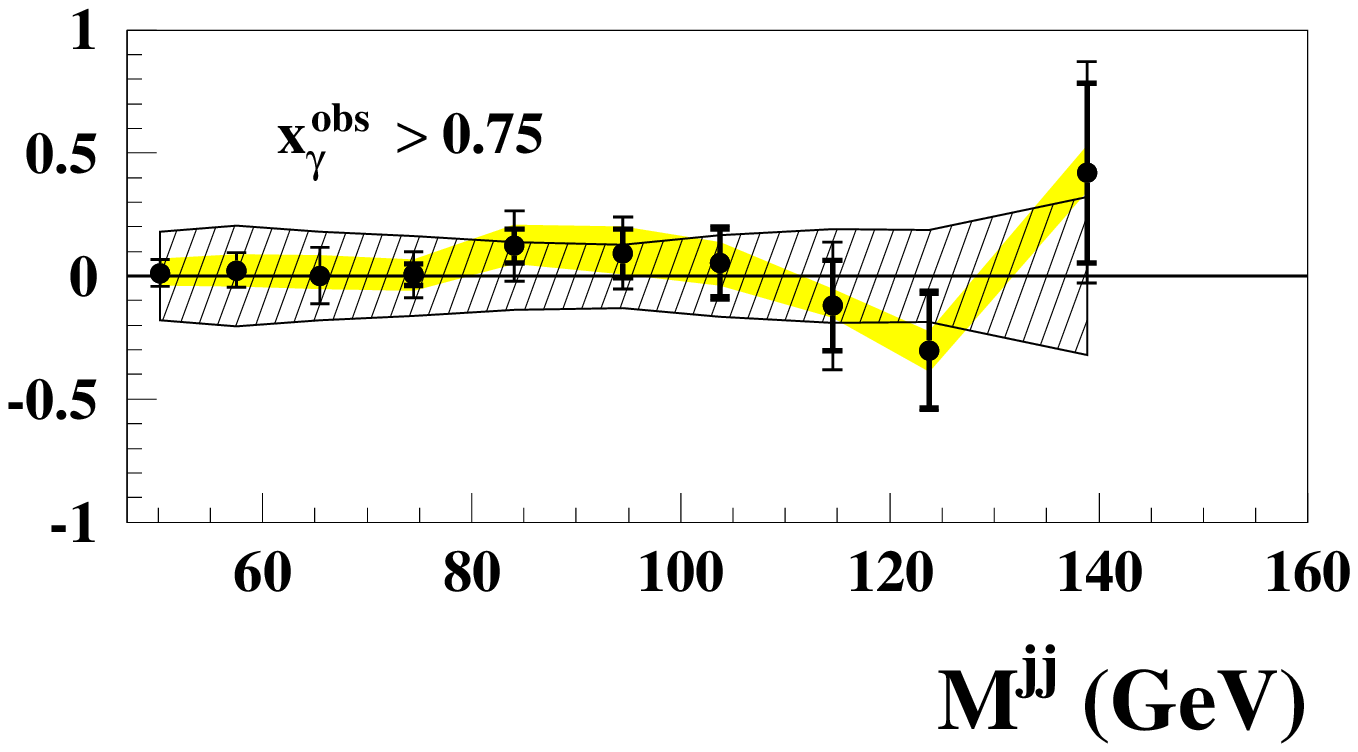,width=15cm}}
\put (12.0,12.5){\small (a)}
\put (12.0,7.75){\small (b)}
\put (12.0,4.00){\small (c)}
\end{picture}
\caption{\label{fig1}
{(a) Dijet cross-section $\smj$; the upper 
data points are for the full region in $\xo$ and the lower points are for
the region $\xo >0.75$. The thick vertical bars represent the statistical
uncertainties of the data, and the thin bars show the statistical and
systematic uncertainties added in quadrature, 
except for that associated with the uncertainty
in the absolute energy scale of the jets (shaded
band).
The LO (dashed line) and NLO (solid line) QCD parton-level calculations
are shown. Both the measurements and calculations of $\smj$,
for $\xo >0.75$, have been multiplied by $0.1$ for clarity of
display; (b) the fractional difference between the measured $\smj$ and the
NLO QCD calculation integrated over the full region in $\xo$; (c) as (b),
but for the region $\xo >0.75$. In (b) and (c), the hatched bands display
the uncertainty of the NLO QCD calculations.}}
\end{figure}

The LO and NLO QCD calculations describe the shape of the measured
$\smj$ well over the entire range of $\mj$ and $\xo$, see Fig.~\ref{fig1}(a). The
shape of this distribution in the calculations is dictated by the $x$ dependence
of the photon and proton PDFs and by the dependence on the 
photon-parton or parton-parton
centre-of-mass energy of the subprocess cross section. The LO QCD
calculation of $\smj$ is $\sim 25\%$ below the data. The inclusion of the NLO
corrections significantly improves the description of the data. Although the data
are still consistently above the calculations, the NLO QCD calculations are
consistent with the data given the present theoretical ($\sim 15\%$) and
experimental ($\sim 10\%$) uncertainties, as shown in Fig.~\ref{fig1}(b).

Since the region $\xo > 0.75$ is dominated by direct processes, in which the
photon behaves as a point-like particle, measurements restricted to that region
in $\xo$ allow a test of the pQCD predictions with less influence from the photon
PDFs. The NLO QCD predictions of $\smj$ have been compared to the data in the region
$\xo >0.75$. In this restricted range, the NLO QCD calculations describe the data well
both in shape and magnitude, as shown in Fig.~\ref{fig1}(c).

Overall, no significant deviation between data and NLO
QCD calculations is observed up to $\mj = 139$~GeV.

\subsection{Dijet angular distribution}
\begin{figure}
\setlength{\unitlength}{1.0cm}
\begin{picture} (18.0,15.0)
\put (-1.0,0.0){\epsfig{figure=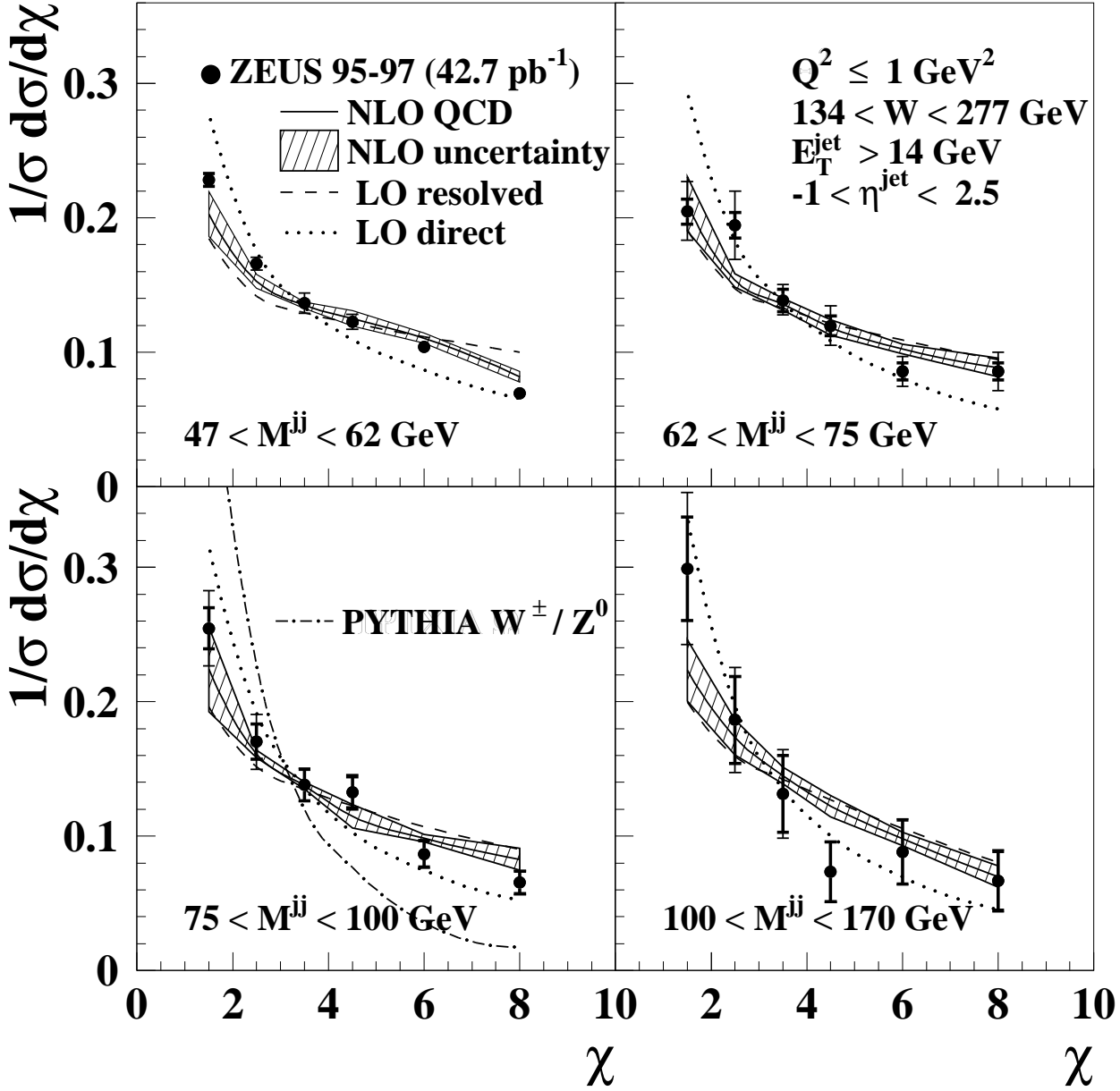,width=18cm}}
\end{picture}
\caption{\label{fig2}
{The normalised dijet cross sections, $\schi$, in the full region of $\xo$,
for four ranges in $\mj$. The thick vertical bars represent the statistical
uncertainties of the data while the thin bars show the statistical and
systematic uncertainties added in quadrature. For comparison, NLO QCD parton-level
calculations are shown as solid lines. The hatched bands display the
uncertainty of the NLO QCD calculations. The LO QCD parton-level calculations
for resolved (dashed lines) and direct (dotted lines) processes are also shown.
The prediction of PYTHIA for the production of $\wp$ and $\z0$ bosons
(dot-dashed line) is also shown in the region $75 < \mj < 100$~GeV.}}
\end{figure}

\begin{figure}
\setlength{\unitlength}{1.0cm}
\begin{picture} (18.0,15.0)
\put (-1.0,0.0){\epsfig{figure=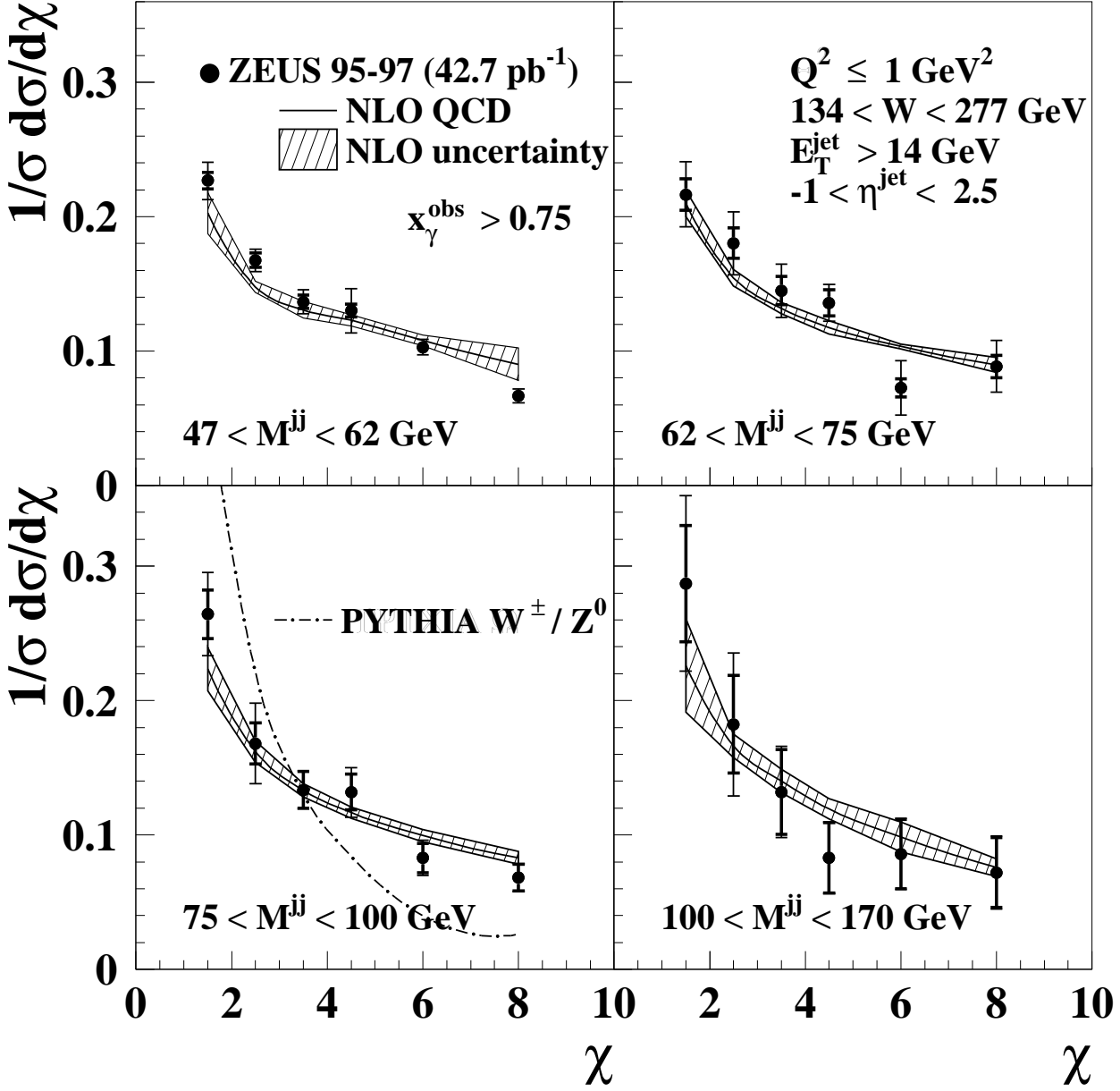,width=18cm}}
\end{picture}
\caption{\label{fig3}
{The normalised dijet cross sections $\schi$, restricted to
$\xo >0.75$, for four ranges in $\mj$. Other details are as described
in the caption to Fig.~\ref{fig2}.}}
\end{figure}

\begin{figure}
\setlength{\unitlength}{1.0cm}
\begin{picture} (18.0,15.0)
\put (-1.0,-1.7){\epsfig{figure=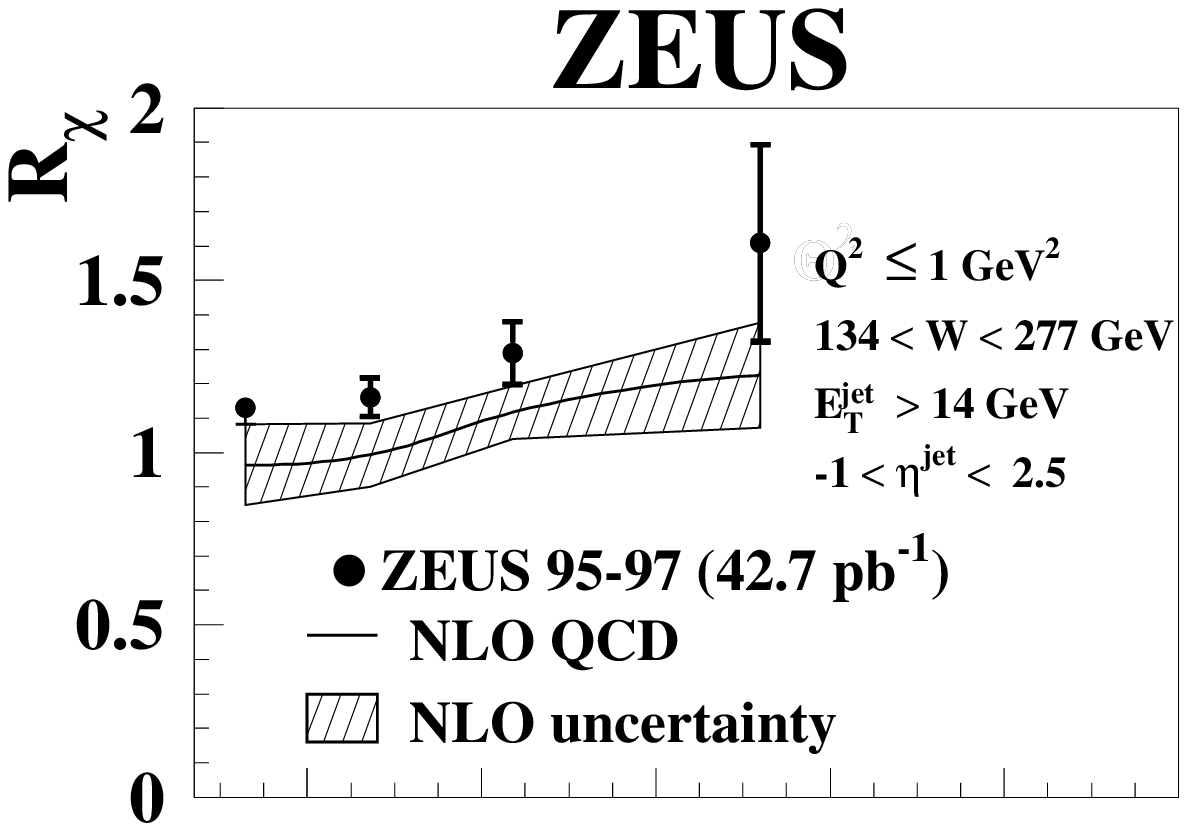,width=18cm}}
\put (2.7,8.0){\epsfig{figure=whitecbg.eps,width=0.5cm,height=0.5cm}}
\put (-1.0,-8.0){\epsfig{figure=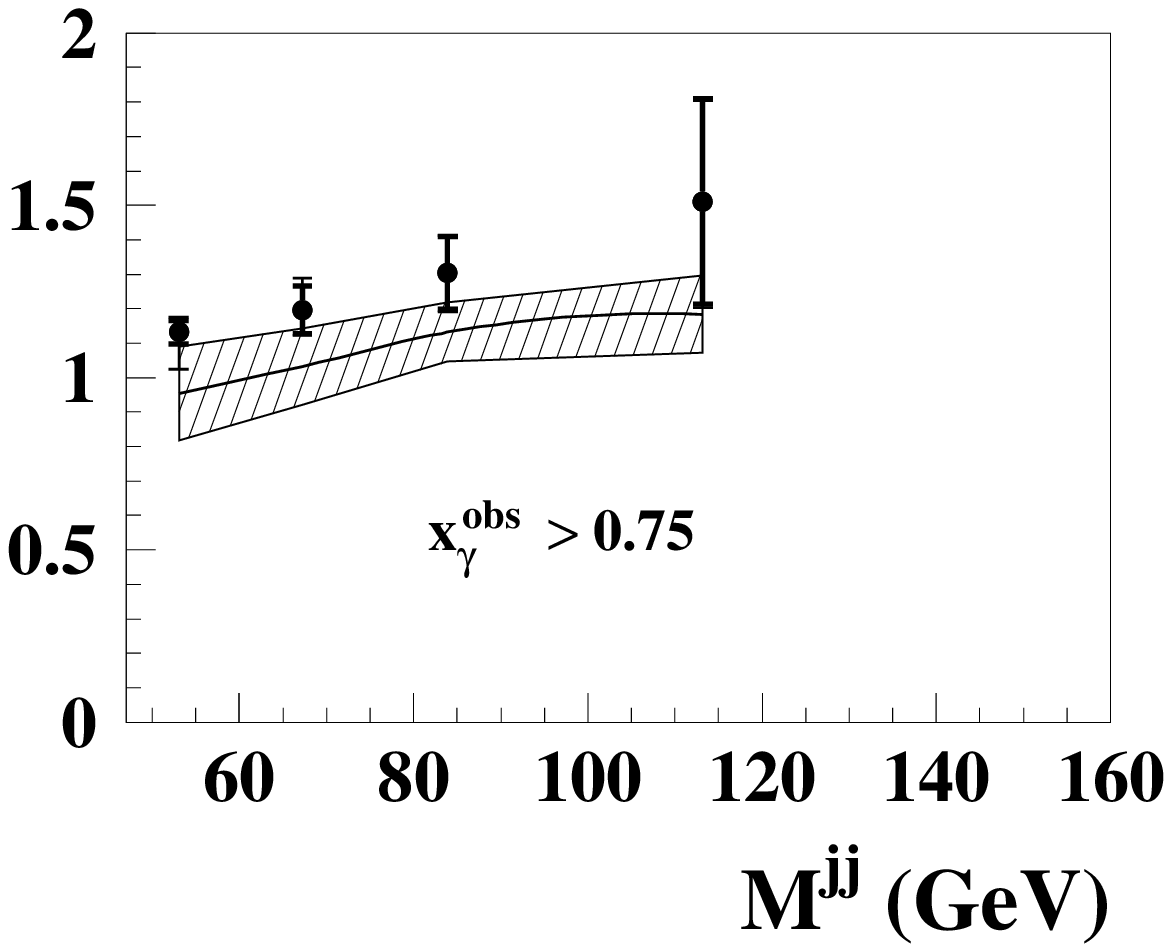,width=18cm}}
\put (11.5,13.5){\small (a)}
\put (11.5,7.0){\small (b)}
\end{picture}
\caption{\label{fig4}
{The cross-section ratio, $\rx = {\rxe}$, as a function of $\mj$:
(a) integrated over the full region in $\xo$; (b) for the region $\xo >0.75$.
The thick vertical bars represent the statistical uncertainties of the data
while the thin bars show the statistical and systematic uncertainties 
added in quadrature. For comparison, the parton-level prediction of NLO QCD
is shown as the solid line; the hatched band displays its uncertainty.}}
\end{figure}

Figure~\ref{fig2} shows that the shape of the NLO QCD calculation is in agreement with
that of the measured $\schi$ in each region of dijet mass studied. An equally good
description is obtained for the measurements restricted to $\xo > 0.75$, as can be
seen in Fig.~\ref{fig3}. The distribution in $\chi$ reflects the spin of the exchanged
particle: it is approximately uniform for two-body processes dominated by gluon
exchange \cite{maxwell} and proportional to $1/(1+\chi)$ for processes mediated by
quark exchange. As an illustration, the LO QCD calculations for resolved and direct
processes are shown in Fig.~\ref{fig2}: the $\chi$ distribution for direct processes,
which are mediated by quark exchange, is steeper than that of resolved processes,
which are dominated by gluon exchange. The measured angular distribution is consistent
with the pQCD description in terms of parton exchange up to the highest $\mj$.

The ratio $\rx\equiv{\sigma(1<\chi<\chi_0)\over\sigma(\chi_0<\chi<9)}$ is
useful to characterise, with a single number, the shape of the angular distribution in
each range of dijet mass \cite{ua1def}. This ratio is shown in Fig.~\ref{fig4},
with the value of $\chi_0$ chosen to be $4$, so that $\rx$ is approximately $1$.
In this ratio, most of the systematic uncertainties cancel out, yielding a 
systematic uncertainty of less than $4\%$ for all $\mj$ ranges considered.
The uncertainties of the theoretical predictions for $\rx$ due to the
choice of photon PDFs and from the choice of renormalisation and factorisation
scales are also reduced. The NLO QCD calculations are consistently below, but
compatible with, the data within the experimental ($\sim 8\%$) and theoretical
($\sim 10\%$) uncertainties.

\subsection{Upper limits on $\z0$, $\wp$ and new heavy-resonance production}

The production of the electroweak gauge bosons $\wp$ and $\z0$ has been studied
in the region $75 < \mj < 100$~GeV for the full range\footnote{Although the
LO contribution to the production of $\wp$ and $\z0$ bosons is given by the
resolved-photon processes (see Section~\ref{secmc}), the resulting $\xo$
distribution peaks at $\xo \sim 0.75$.} in $\xo$. The $\schi$ distribution is not
consistent with a large contribution from $\wp$ and $\z0$ decays, as can be seen
from Fig.~\ref{fig2}, which shows that QCD processes are dominant. Simulations of
both the QCD (non-resonant) dijet photoproduction background and the $\z0$ signal
were used to find values that optimise the observation of a $\z0$ signal relative
to the background for the upper cut on the measured value of $\cost$,
$\cost_{\rm cut}$, and the dijet-mass window. The background expected from QCD
processes was estimated by using the NLO QCD predictions with the normalisation
obtained from a fit to the measured $\smj$ in the region $47 < \mj < 75$~GeV. 
Since there is no evidence for a signal, an upper limit on the cross section
for $\z0$ production has been derived. In the dijet-mass window of $91\pm 9$~GeV
and for $\cost < 0.6$, 230 events were observed while 223 events were
expected from QCD processes. The acceptance for $\z0$ production after all the
selection cuts was $14.6\%$. The resulting $95\%$ C.L. upper limit in the kinematic
region $\q2\; \leq\; 1$~\g2\ is
$$\sigma_{e^+ p \rightarrow e^+ \z0 X} < 5.9~{\rm pb}.$$
In deriving the upper limit, the two different photon PDFs mentioned in
Section~\ref{secnlo} were used; the one that gave the most conservative limit was
chosen. This upper limit on $\z0$ production is the first obtained at HERA. The
Standard Model expectation for the process $e^+ p \rightarrow e^+ \z0 X$ integrated
over all $\q2$ is $0.3$~pb~\cite{baur,wzeus}. The same procedure was applied for
$\wp$ production and an upper limit of
$\sigma_{e^+ p \rightarrow e^+ \wp X} < 7.4~{\rm pb}$ at $95\%$ C.L. was obtained.
This limit is weaker than that already obtained by ZEUS using the $\wp$ leptonic
decays \cite{wzeus}.
The Standard Model expectation for the cross section of the process
$e^+ p \rightarrow e^+ \wp X$ is $0.95$~pb~\cite{baur,wzeus}.

The increase of the $\gp$ centre-of-mass energy by an order of magnitude with
respect to fixed-target experiments \cite{oldgp} allows a search for generic heavy
resonances, denoted by $\cal{P}$, with masses above $60$~GeV in the
reaction $\gp \rightarrow {\cal{P}} + {\rm X} \rightarrow {\rm jet} + {\rm jet} + {\rm X}$.
While searches for such resonances formed by $q\bar{q}$ annihilation, such as
$Z^{\prime}$, have been performed \cite{ua2search} with similar sensitivity
in this mass range, there are models such as Technicolour \cite{techno} which
predict resonances\footnote{Some Technicolour models \cite{cornet} predict the
existence of light colour-octet, isospin-singlet pseudo-Goldstone bosons
that would decay into two gluons.} produced preferentially by photon-gluon fusion.
Such particles would not have been observed in earlier searches.

\begin{figure}
\setlength{\unitlength}{1.0cm}
\begin{picture} (18.0,15.0)
\put (-3.5,-12.0){\epsfig{figure=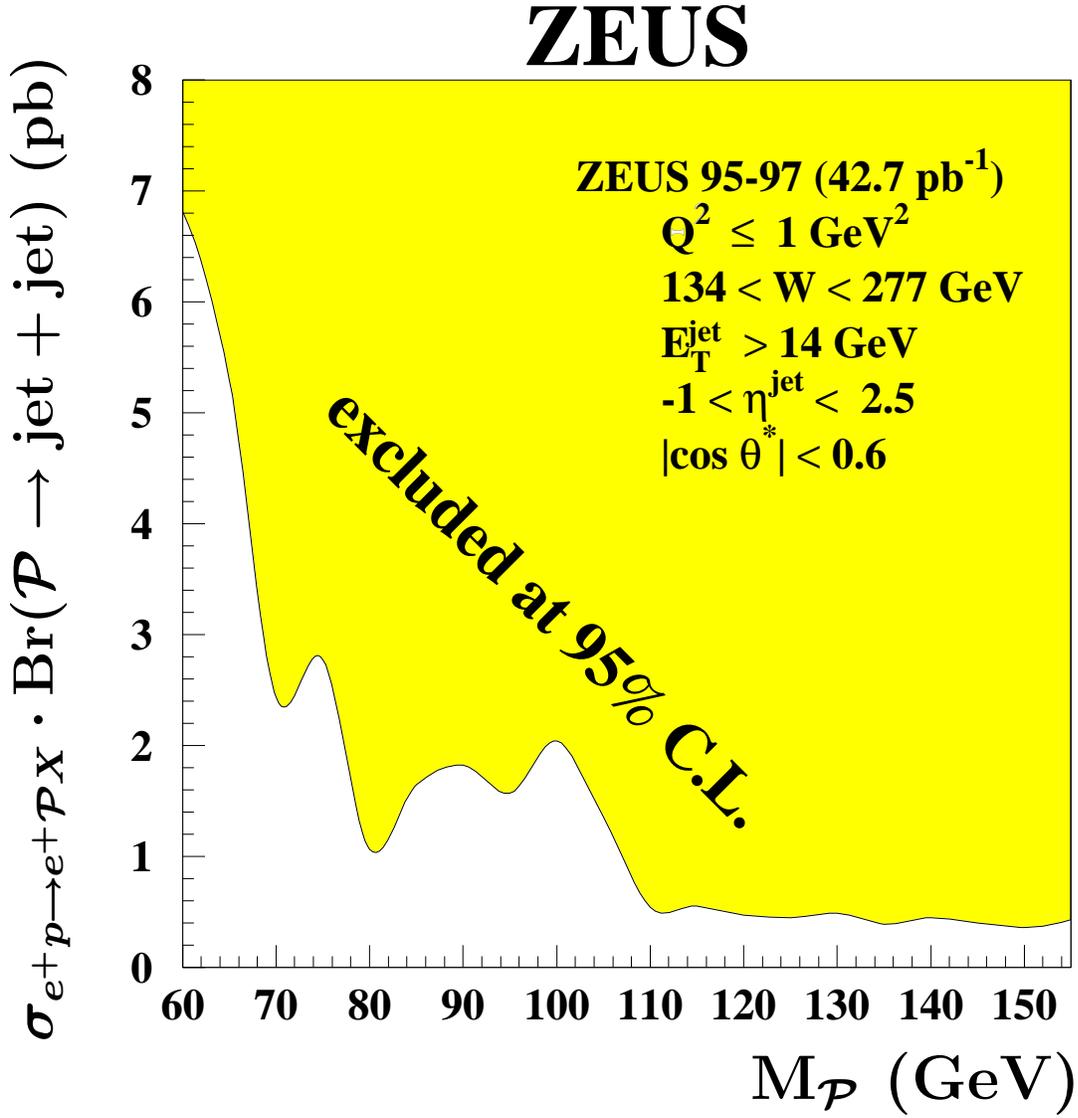,width=23cm}}
\end{picture}
\caption{\label{fig5}
{Upper limit at $95\%$~C.L. on the cross section times branching ratio for the production of
new heavy resonances decaying into two jets,
$\sigma_{e^+p\rightarrow e^+{\cal{P}}X} \cdot {\rm Br}({\cal{P}} \rightarrow
{\rm jet}+{\rm jet})$, as a function of the resonance mass, $M_{\cal{P}}$.}}
\end{figure}

A search for the photoproduction of a new heavy resonance decaying into two jets
has been performed in the mass range $60 < M_{\cal P}< 155$~GeV. The method
described above has also been used in this search to select the $\mj$ window for
$\cost <0.6$. In this case, the background expected from QCD processes has been
estimated by using the NLO QCD predictions with the normalisation obtained from a fit
to the measured $\smj$ in the region of $\mj$ values well below that of the signal.
Since there is no evidence for a signal, an upper limit on the cross section times the
branching ratio into two jets for $\cal{P}$ production,
$\sigma_{e^+p\rightarrow e^+{\cal{P}}X} \cdot {\rm Br}({\cal{P}} \rightarrow
{\rm jet}+{\rm jet})$,
has been derived. The limit refers to the kinematic range defined by
$\q2 \; \leq \; 1$~\g2, \wr , $\etjet > 14$~GeV, $\etar$ and $\cost <0.6$. The
restriction to $\cost <0.6$ reduces the background from QCD-induced processes
and the dependence on the spin and decay angular distribution of the heavy resonance
(see Section 5). The upper limit on
$\sigma_{e^+p\rightarrow e^+{\cal{P}}X} \cdot {\rm Br}({\cal{P}} \rightarrow
{\rm jet}+{\rm jet})$
at $95\%$~C.L. is shown in Fig.~\ref{fig5} as a function of $M_{\cal{P}}$.

\section{Summary}
\label{secsumm}

Measurements of differential cross sections for dijet photoproduction have been made
in $e^+p$ collisions at a centre-of-mass energy of $300$~GeV using $42.7$~\pb1\ of data
collected with the ZEUS detector at HERA. The dijet cross sections refer
to jets identified with the $\kt$ cluster algorithm in the longitudinally invariant
inclusive mode which were selected with $\etjet > 14$~GeV and $\etar$. The
measurements were made in the kinematic region defined by $\q2 \; \leq \; 1$~\g2\ and
\wr . The dijet differential cross section as a function of $\mj$ has been measured for
$\mj >47$~GeV and $\cost<0.8$. Values of $\mj$ up to $\sim 160$~GeV are
accessible with the present data. The NLO QCD calculations give a good description
of the shape of the measured differential cross-section $\smj$. 

The dijet cross section as a function of $\chi=(1+\cost)/(1-\cost)$ has been measured in
several $\mj$ ranges. The dependence of the distribution $\schi$ and of the ratio
$\rx\equiv{\rxe}$ on the dijet mass are found to be consistent with the predictions
of NLO QCD in the $\mj$ range studied. The observed agreement in shape between data
and the QCD calculations from $\mj = 50$~GeV up to $\mj = 139$~GeV confirms the
validity of the pQCD description of photon-parton interactions down to distances of
approximately $1.4 \cdot 10^{-3}$~fm.

The dijet events in the region $75< \mj < 100$~GeV have been used
to derive an upper limit on the cross section for $\z0$ production of 
$\sigma_{e^+ p \rightarrow e^+ \z0 X} < 5.9$~pb at $95\%$~C.L. for
$Q^2 \; \leq \; 1$~GeV$^2$. Upper limits on the photoproduction of new heavy
resonances decaying into two jets have been presented for $60 < M_{\cal{P}}< 155$~GeV.

\vspace{0.5cm}
\noindent {\Large\bf Acknowledgements}
\vspace{0.3cm}

We thank the DESY Directorate for their strong support and encouragement.
The remarkable achievements of the HERA machine group were essential for
the successful completion of this work and are greatly appreciated. We
thank M. Klasen for valuable discussions and help in running
his program for NLO QCD calculations.


\newpage
\clearpage
\parskip 0mm

\newpage
\clearpage
\parskip 0mm

\newpage
\clearpage
\parskip 0mm

\newpage
\clearpage
\parskip 0mm

\newpage
\clearpage
\parskip 0mm

\end{document}